\documentclass[lettersize,journal]{IEEEtran}
\usepackage{amsmath,amsfonts}
\usepackage{algorithmic}
\usepackage{algorithm}
\usepackage{array}
\usepackage[caption=false,font=normalsize,labelfont=sf,textfont=sf]{subfig}
\usepackage{textcomp}
\usepackage{stfloats}
\usepackage{url}
\usepackage{verbatim}
\usepackage{graphicx}
\usepackage{cite}

\usepackage{bm}
\usepackage[colorlinks,linkcolor=blue,anchorcolor=blue,citecolor=blue]{hyperref}    
\begin{document}

\title{VQ-DeepVSC: A Dual-Stage Vector Quantization Framework for Video Semantic Communication}

\author{Yongyi Miao\textsuperscript{+}, Zhongdang Li\textsuperscript{+}, Yang Wang, Die Hu\textsuperscript{*}, Jun Yan, and Youfang Wang
\thanks{Yongyi Miao, Zhongdang Li, Yang Wang, Die Hu, Jun Yan, and Youfang Wang are with the School of Information Science and Technology, Fudan University, Shanghai 200433, China.}
}



\maketitle
\begin{abstract}
In response to the rapid growth of global video traffic and the limitations of traditional wireless transmission systems, we propose a novel dual-stage vector quantization framework, VQ-DeepVSC, tailored to enhance video transmission over wireless channels. 
In the first stage, we design the adaptive key-frame extractor and interpolator, deployed respectively at the transmitter and receiver, which intelligently select key frames to minimize inter-frame redundancy and mitigate the \emph{cliff-effect} under challenging channel conditions. 
In the second stage, we propose the semantic vector quantization encoder and decoder, placed respectively at the transmitter and receiver, which efficiently compress key frames using advanced indexing and spatial normalization modules to reduce redundancy. 
Additionally, we propose adjustable index selection and recovery modules, enhancing compression efficiency and enabling flexible compression ratio adjustment.
Compared to the joint source-channel coding (JSCC) framework, the proposed framework exhibits superior compatibility with current digital communication systems.
Experimental results demonstrate that VQ-DeepVSC achieves substantial improvements in both Multi-Scale Structural Similarity (MS-SSIM) and Learned Perceptual Image Patch Similarity (LPIPS) metrics than the H.265 standard, particularly under low channel signal-to-noise ratio (SNR) or multi-path channels, highlighting the significantly enhanced transmission capabilities of our approach.
\end{abstract}

\begin{IEEEkeywords}
Semantic Communication, Video Transmission, Vector Quantization, Multipath Fading Channel, Deep Learning.
\end{IEEEkeywords}

\section{Introduction}
\IEEEPARstart{W}{ith} the rapid advancement of information technology, global data traffic, especially video traffic, has grown exponentially, now constituting the predominant component of internet data traffic \cite{al_fuqaha_internet_2015, zhang_6g_2019}. Despite enhancements in traditional wireless video transmission systems that focus on optimizing bit error rates (BER) \cite{wiegand_overview_2003, sullivan_overview_2012}, challenges persist in ensuring high-quality transmission. These systems primarily emphasize compression efficiency but often fall short in addressing the semantic understanding and adaptability required for dynamic network conditions. Even with the adoption of the latest H.265 technology, the impact of the \emph{cliff-effect} remains unresolved. This effect describes a significant decline in video transmission quality when the channel signal-to-noise ratio (SNR) drops below a critical threshold.

\subsection{Prior Work}
The rapid development of deep learning (DL) technology has propelled semantic communication into a crucial role in the next generation of communication technologies, demonstrating significant potential and advantages \cite{shi_semantic_2021, luo_semantic_2022, yang_semantic_2023}. Semantic communication, grounded in understanding before transmitting, involves extracting semantic information \cite{xu_wireless_2022, hu_robust_2022}. It enables profound compression by analyzing and distilling the essence of original content, proving highly adaptable and extensible across diverse domains, including text \cite{zhou_semantic_2022}, speech \cite{tong_federated_2021, weng_semantic_2021}, images \cite{xu_wireless_2022-1, hu_robust_2023, bourtsoulatze_deep_2019}, and video \cite{tung_deepwive_2022}. Thus, it heralds a new era of intelligent and context-aware communication systems.

Despite considerable focus on the application of deep learning to video compression~\cite{ma_image_2020, yang_learning_2020, lu_end--end_2021, hu_fvc_2021}, the advent of semantic-based deep learning to enhance wireless video transmission was marked by the introduction of DeepWiVe by Tung et al. in 2022 \cite{tung_deepwive_2022}. This pioneering approach presented an integrated solution for video compression, channel coding, and modulation through an end-to-end joint source-channel coding (JSCC) framework. Building upon this foundation, Zhang et al. further innovated by incorporating dual optical flow estimation for video transmission \cite{zhang_deep_2023}. Dong et al. introduced Rosefinch, a semantic communication model for live media streaming, video conferencing, and low-rate image transmission~\cite{dong_demo_2022}. With these developments, JSCC framework has become the main research paradigm for video semantic communication (VSC) systems.

Notably, Gong et al. \cite{gong_adaptive_2023} introduced an adaptive bit rate VSC system, modulating the bit rate based on network conditions. Niu et al. \cite{niu_deep_2023} added a channel and spatial attention mechanism to their VSC framework, enhancing adaptability. Bao et al. \cite{bao_mdvscwireless_2023} proposed a model division VSC scheme to extract shared semantic features and counteract noisy channels. Liang et al. \cite{liang_vista_2023} developed the VISTA framework, using semantic location graphs to manage dynamic objects and adapt to changing channel conditions. 
Specialized systems have also been developed for specific scenarios. Jiang et al. \cite{jiang_wireless_2023} created a scalable video coding network for video conferencing, minimizing transmission resources. Liu et al. \cite{liu_efficient_2023} introduced a federated learning-enhanced vehicle semantic communication framework, optimizing semantic extraction and resource allocation while maintaining data privacy. Wang et al. \cite{wang_wireless_2023} employed a nonlinear transformation and conditional coding architecture to optimize the balance between transmission rate and distortion, underpinned by perceptual quality metrics. 
\subsection{Motivation and Contributions}
While JSCC-based VSC systems have theoretically shown promise in reducing distortion and enhancing transmission efficiency, these systems map source data directly to channel symbols, causing constellation points to appear anywhere on the constellation diagram, which deviates from the design of current digital communication systems. 
In addition, most studies simulate channels based on additive white Gaussian noise (AWGN) or other simplistic models. Although the AWGN model is useful for controlled experiments, it fails to capture the complexities of real-world wireless communications, such as multipath effects, signal decay, and the ever-fluctuating nature of channel conditions. 
Therefore, the robustness and suitability of JSCC methodologies in complex, dynamic wireless environments require thorough validation.

To make semantic communication more compatible with digital communication systems, semantic communication systems for image transmission based on discrete latent spaces have been designed \cite{hu_robust_2022} \cite{fu_vector_2023} so that semantic features can be quantized into feature indices through the latent embedding space, thereby converting them into bitstreams. Subsequently, these bitstreams can be directly mapped into symbols for transmission using existing constellation mapping schemes.

The vector quantization (VQ) semantic communication system designed by Hu et al.~\cite{hu_robust_2022}, Masked VQ-VAE, reduces transmitted information by quantizing and transmitting only feature indices. However, due to the limitations of VQ-VAE in image reconstruction~\cite{VQVAE}, it addresses downstream tasks solely by transmitting task-relevant key feature indices within the latent embedding space. In addition, VQ-DeepSC, designed by Fu et al.~\cite{fu_vector_2023}, utilizes a U-Net structure~\cite{Unet} to extract features at different scales and the input image is mapped to single-channel indices at each scale. However, this method necessitates multiple latent embedding spaces of different sizes, leading to excessive storage requirements at both the transmitter and receiver ends. Furthermore, the information bottleneck problem within the U-Net structure limits its ability to capture remote global contextual information, thereby reducing its effectiveness in handling high-resolution images. Additionally, the quantization operator is lossy, and similar patches are often embedded with the same index, leading to the creation of pseudo-shadows and discontinuities in generated images. 

In addition to being limited by the image reconstruction capabilities of current semantic communication network structures, extending VQ-based semantic communication systems from image to video transmission is challenging. This is due to the unique demands of video transmission, such as the need to further reduce temporal redundancy.

In this paper, we firstly propose a dual-stage vector quantization-based video semantic communication system named VQ-DeepVSC. To enhance the physical interpretability of the system, VQ-DeepVSC is divided into two distinct stages to improve transmission efficiency, each specifically aimed at reducing redundancy in the temporal and spatial dimensions, respectively. 

In the first stage, we design an adaptive key-frame extraction and interpolation (AKEI) module to reduce temporal redundancy, transforming video transmission into a key-frame transmission task. Meanwhile, to mitigate the \emph{cliff-effect} arising from the constrained channel coding capacity in separated source-channel coding systems, the proposed AKEI estimates frame importance and models the relationship between SNR and key-frame rate, prioritizing the retransmission of critical key frames under poor channel conditions (e.g., when SNR is  low). At the receiving end, an optical flow estimation-based interpolation algorithm restores the video to its original frame rate, enhancing robustness and stability in challenging channel conditions. 
In the second stage, we leverage a multi-channel spatially conditioned vector quantization (MSVQ) for the ultimate compression of key frames. Compared to existing methodologies such as Masked VQ-VAE~\cite{hu_robust_2022} and VQ-DeepSC~\cite{fu_vector_2023}, MSVQ utilizes a shared latent embedding space for multi-channel index mapping~\cite{MOC-RVQ}, supporting high-resolution video transmission and promoting diversity without increasing latent space size. Spatial conditional normalization in the MSVQ decoder mitigates quantization artifacts, ensuring high-quality image reconstruction. Additionally, key-frame indices are further compressed using adjustable index selection and recovery algorithms, allowing flexible compression rates. The main contributions of this paper are summarized as follows:
\begin{itemize}
    \item We are the first to use DL-based VQ methods for video transmission. Compared to JSCC-based VSC systems, the VQ-DeepVSC adapts better to the variable conditions of real wireless channels and is more compatible with existing communication systems. The proposed MSVQ is designed to minimize intra-frame redundancies using multi-channel indexing and spatial conditional normalization, achieving a high compression degree while maintaining frame quality.
    \item We design the AKEI module, a revolutionary framework that intelligently selects key frames based on content and wireless channel quality, thereby optimizing the semantic compression process, significantly reducing inter-frame redundancies, and effectively tackling the \emph{cliff-effect} in video transmission.
    \item We propose adjustable index selection and recovery algorithms that reduce redundancy between key frames by transmitting only indices significantly different from their predecessors, as determined by a preset threshold. This approach enhances compression efficiency, minimizes compression loss, and allows for more flexible adjustment of compression rates.
    \item Extensive simulation experiments have validated that the proposed VQ-DeepVSC achieves superior reconstruction quality at equivalent compression rates compared to other methods. What's more, our system demonstrates exceptional robustness, maintaining high-performance transmission efficiency across various channel conditions.
\end{itemize}

\subsection{Organization}
The structure of this paper is as follows: Section~\ref{section2} proposes the framework of the VQ-DeepVSC system. Section~\ref{section3} elaborates on the proposed AKEI. Section~\ref{section4} designs the implementation of the MSVQ. Section~\ref{section5} explores the strategy for reducing redundancy between key frames. Section~\ref{section6} presents experimental results that validate the performance of the proposed VQ-DeepVSC system. Finally, Section~\ref{section7} concludes the paper.

\emph{Notations}: Superscripts $^T$ stands for transpose. $\mathbf{I}_{\mathit{M}\times \mathit{N}}$ represents the $\mathit{M}\times \mathit{N}$ identity matrix and $\mathbf{0}_{M\times N}$ denotes the $M\times N$ all-zero matrix. For a matrix \(\mathbf{A}\), \([\mathbf{A}]_{\mathit{i},:}\) denotes the $i$-th row of \(\mathbf{A}\).
For a vector $\mathbf{a}$, $||\mathbf{a}||_2$ denotes the Euclidean norm, $\mathbf{a}(m)$ denotes the $ m $-th elements of $\mathbf{a}$, and $\mathbf{a}(\mathit{m}:\mathit{n})$ represents taking the $\mathit{m}$-th to $\mathit{n}$-th elements from vector $ \mathbf{a} $. 
For a scalar $x$, \(\left\lfloor x \right\rfloor\) denotes the nearest integer smaller than or equal to $x$.
For operations, \(\langle \cdot, \cdot \rangle\) represents the inner product, $\otimes$ indicates the batch-wise matrix multiplication operation, and $\odot$ denotes the Hadamard product.
Finally, \(\mathbb{R}^{m \times n}\) and \(\mathbb{Z}^{m \times n}\) denote the spaces of \(m \times n\) real and integer matrices, respectively.

\section{Framework of VQ-based Semantic Communications for Video Transmission} \label{section2}

In this section, we introduce a novel framework for semantic communication systems specifically tailored for video transmission. This framework, based on VQ, 
is illustrated in Fig.~\ref{fig:whole_figure}. The transmitter consists of an adaptive key-frame extractor, a semantic vector quantization encoder, an adjustable index selector, channel coding and modulation, and orthogonal frequency division multiplexing (OFDM) modulation. The receiver comprises OFDM demodulation, channel estimation and channel equalization, a demodulation and channel decoder, an adjustable index restorer, a semantic vector quantization decoder, and an adaptive key-frame interpolator.
\subsection{Transmitter}
\begin{figure*}[t]
    \centering
    \includegraphics[width=0.85\textwidth]{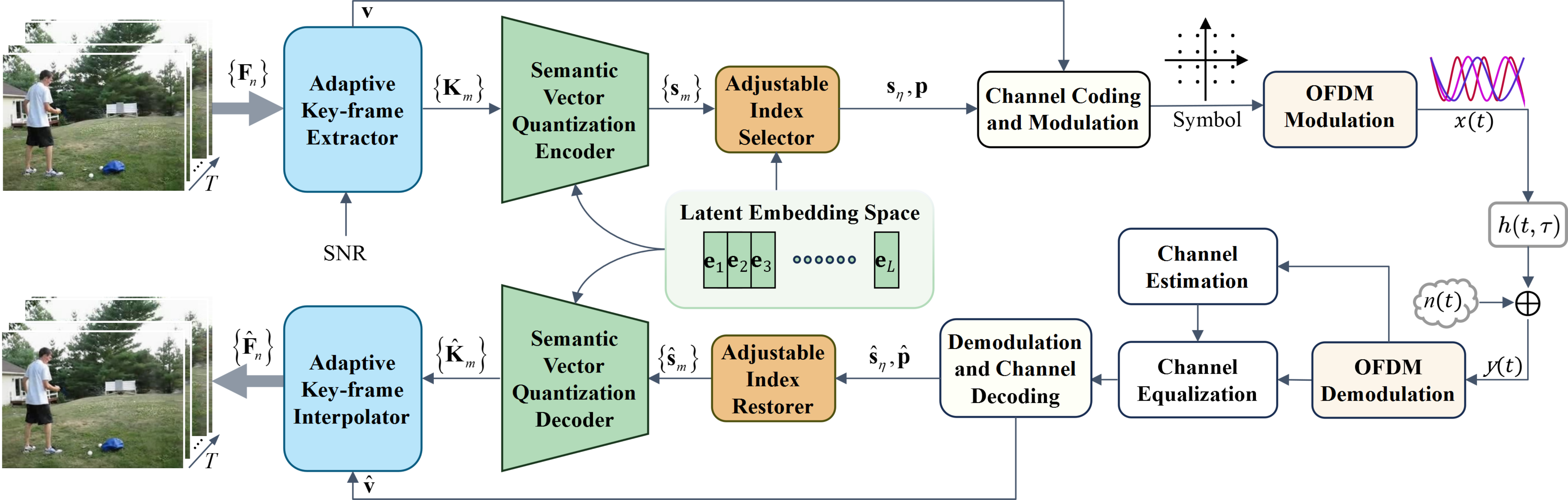}%
    \caption{The overall system architecture of the proposed VQ-DeepVSC.}
    \label{fig:whole_figure}
    \vspace{-10pt}
\end{figure*}

As shown in Fig.~\ref{fig:whole_figure}, the video transmitter processes a sequence of video data, which spans an arbitrary duration $T$ and is composed of \( N \) individual frames. Each frame \(\mathbf{F}_n \in \mathbb{R}^{W_\text{F} \times H_{\text{F}} \times C_\text{F}}\) with $n = 1, \ldots, N$ is an RGB image with the width of \( W_{\text{F}} \), the height of \( H_{\text{F}} \) , and the \( C_{\text{F}} = 3\) channels.

To avoid the \emph{cliff-effect}, a feasible method is to retransmit video frames multiple times. However, this approach will undoubtedly increase the amount of transmitted data, greatly affecting the compression effect. To ensure that the compression ratio remains unchanged during retransmission, we propose an adaptive key-frame extractor that assesses the importance of each frame based on channel quality (such as SNR) and content changes, and extracts the key frames. Only the key frames are transmitted and retransmitted under poor channel conditions, and denote the $m$-th key frame as $\mathbf{K}_m \in \mathbb{R}^{W_\text{F} \times H_\text{F} \times C_\text{F}}$, where $m = 1, \ldots, M$, and $M$ represents the number of key frames, and we use the vector $\mathbf{v} \in \mathbb{Z}^{N \times 1}$ to record the positions of the key frames. Specifically, if the \(\mathit{n}\)-th frame is the key frame, we set $\mathbf{v}(n)=1$ , otherwise we set $\mathbf{v}(n)=0$, where $n = 1, \ldots, N$. The specific details of this process and its implications will be elaborated in Section \ref{sec:decoder_interpolator}.

We define the ratio of key frames to all frames $\rho$ as:
\begin{align}
\rho = \frac{M}{N}, \label{rou}
\end{align}
which can be adaptively adjusted according to the channel conditions.

After extracting key frames, a semantic vector quantization encoder is utilized for feature extraction and indexing. Here, we utilize a latent embedding space which is denoted by \(\mathbf{E} = \{\mathbf{e}_{1}, \ldots, \mathbf{e}_{L}\} \in \mathbb{R}^{L \times d}\), where \( L \) denotes the size of \(\mathbf{E}\), \( d \) is the dimensionality of each latent embedding vector \(\mathbf{e}_{l}\) in \(\mathbf{E}\), and \( l = 1, \ldots, L \). Space \(\mathbf{E}\) is trained during the training phase and is shared between the transmitter and receiver. Following feature extraction by the semantic vector quantization encoder, \(\mathit{M}\) index sequences \(\{\mathbf{s}_m\}_{m=1}^M\) are generated, where \(\mathbf{s}_m\) represents the index sequence of the \(m\)-th keyframe corresponding to the quantized vector in the \(\mathbf{E}\) space, \( m = 1, \ldots, M \). Details will be provided in Section~\ref{section4}. 

To further reduce redundancy, we propose an adjustable index selector that performs the final screening of the index sequence. The outputs from the adjustable index selector are the resulting indices \(\mathbf{s}_{\eta}\) and the position sequence \(\mathbf{p}\). The specific details will be provided in Section \ref{section5}.

After the aforementioned modules, the video data \( \{\mathbf{F}_n\}_{n=1}^N \) can ultimately be encoded into a bitstream \( \mathbf{b} \), which is composed of the following three parts, i.e.:
\begin{align}
    \mathbf{b} = [\mathbf{b}_\mathbf{s}, \mathbf{b}_\mathbf{p},\mathbf{b}_\mathbf{v}],
\end{align}
where \(\mathbf{b}_\mathbf{s}\), \(\mathbf{b}_\mathbf{p}\) and \(\mathbf{b}_\mathbf{v}\) correspond to the bitstreams of \(\mathbf{s}_{\eta}\), \(\mathbf{p}\) and \(\mathbf{v}\), respectively.

After obtaining $\mathbf{b}$, operations identical to those used in traditional communication systems can be employed, namely encoding and modulating the bitstream for transmission. To combat multipath fading in the channel, OFDM technology is also utilized here. As can be seen from the figure, the semantic communication system we propose is fully compatible with existing communication systems.

\subsection{Receiver}

The received time-domain signal at the receiver can be expressed as:
\begin{align}
y(t) = h(t,\tau) * x(t) + n(t),
\end{align}
where \( h(t, \tau) \) represents the multi-path time-varying channel impulse response, \( n(t) \) denotes the additive white Gaussian noise, and \( * \) signifies convolution.

After applying OFDM demodulation, channel equalization, demodulation, and channel decoding, we can 
finally obtain the estimates of \(\mathbf{s}_{\eta}\), \({\mathbf{p}}\) and ${\mathbf{v}}$, i.e., \(\hat{\mathbf{s}}_{\eta}\), \(\hat{\mathbf{p}}\) and ${\hat{\mathbf{v}}}$, respectively.

Subsequently, the adjustable index restorer is employed based on \(\hat{\mathbf{s}}_{\eta}\) and \(\hat{\mathbf{p}}\), enabling us to retrieve $\{\hat{\mathbf{s}}_m\}_{m=1}^{M}$.
Given \( \hat{\mathbf{v}} \), the semantic vector quantization decoder is then used to recover the key frames, i.e., to obtain \( \{\hat{\mathbf{K}}_m\}_{m=1}^{M} \). Finally, given \( \{\hat{\mathbf{K}}_m\}_{m=1}^{M} \) and \( \hat{\mathbf{v}} \), we use an adaptive key-frame interpolator to recover all video frames \( \{\hat{\mathbf{F}}_n\}_{n=1}^{N} \).

\section{The Proposed AKEI}\label{section3}
\label{sec:decoder_interpolator}
Many frames within a video sequence are often similar, containing redundant information. Therefore, the transmission of data can be reduced by only transmitting a subset of frames and then performing frame interpolation at the receiver. These transmitted frames are known as key frames. Here, we propose AKEI to achieve the extraction of key frames and the subsequent recovery of all frames.

AKEI aims to reduce inter-frame redundancies in video transmission and address \emph{cliff-effect}. AKEI consists of two primary components: the adaptive key-frame extractor, located on the transmitter, and the adaptive key-frame interpolator, situated on the receiver. 

When the SNR decreases, the quality of video transmission may suddenly drop at a certain point, a phenomenon known as the \emph{cliff-effect}. An effective method to address the \emph{cliff-effect} is to perform multiple retransmissions of the video, but this can lead to a significant decrease in compression capability of the system. To solve the \emph{cliff-effect} without increasing the compression rate, we employ an adaptive key-frame extractor to dynamically select key frames for transmission and perform multiple retransmissions under low SNRs. The adaptive key-frame extractor dynamically selects key frames and generates a vector to record the positions of these key frames within the video sequence. This selection process is adaptively adjusted based on content variation between video frames and the quality of the multi-path channel. 

Conversely, the adaptive key-frame interpolator utilizes the received key frames to reconstruct the full video sequence. To facilitate practical deployment, AKEI is designed with an asymmetric structure, featuring a simpler structure at the receiver.

\subsection{Adaptive Key-frame Extractor}\label{AKE}

\begin{figure}[!t]
\centering
\includegraphics[width=0.42\textwidth]{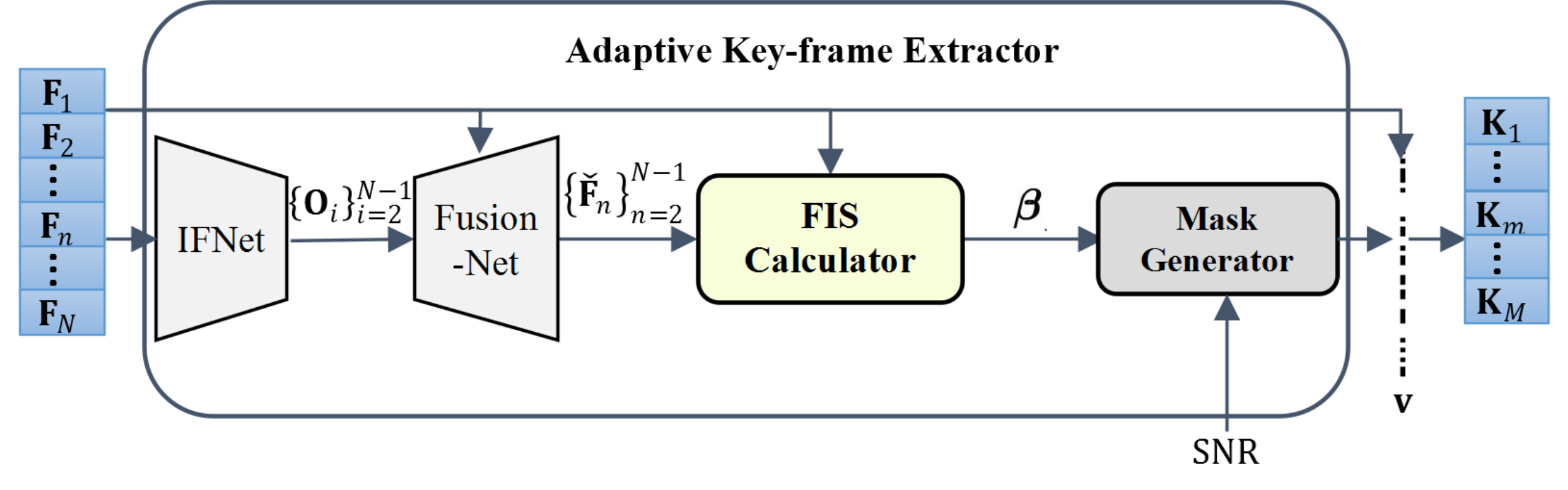}
\vspace{-10pt}
\caption{The structure of the adaptive key-frame extractor.}
\label{fig:AK_extractor}
\vspace{-15pt}
\end{figure}

As illustrated in Fig.~\ref{fig:AK_extractor}, the input of the adaptive key-frame extractor is \(\{\mathbf{F}_n\}_{n=1}^{N}\). And the output is the key frames \(\{\mathbf{K}_m\}_{m=1}^{M}\). Initially, an incremental flow network (IFNet) is used to estimate the optical flow of frames and outputs \(\{\mathbf{O}_i\}_{i=2}^{N-1}\). Each optical flow map \(\mathbf{O}_i \in \mathbb{R}^{C_\text{O} \times H_\text{O} \times W_\text{O}}\) denotes the optical flow corresponding to frame \(\mathbf{F}_i\), where $C_\text{O}$, $H_\text{O}$, and $W_\text{O}$ represent the number of channels, height, and width of \(\mathbf{O}_i\), respectively. These dimensions are consistent across all \(\mathbf{O}_i\).
By default, we assume that $\mathbf{F}_1$ and $\mathbf{F}_N$ are the key frames, i.e., we set $\mathbf{v}(1) = 1$ and $\mathbf{v}(N) = 1$. Therefore, the IFNet only computes the optical flow for the intermediate $(N-2)$ frames. Unlike dual optical flow networks that provide two optical flow estimations for each frame to represent the content changes relative to the preceding and succeeding frames, our approach only outputs one optical flow estimation for each frame, indicating the content changes relative to the surrounding frames. 

Suppose that the frame to be reconstructed is \( \mathbf{F}_i \), where $i = 2, \ldots, N-1$, the FusionNet utilizes its preceding frame \( \mathbf{F}_{i-1} \) and succeeding frame \( \mathbf{F}_{i+1} \), as well as the optical flow estimation \( \mathbf{O}_{i} \) of \(\mathbf{F}_i\), to reconstruct \(\mathbf{F}_i\). Here, we denote the reconstruction result by \( \mathbf{\check{F}}_i \).

Following FusionNet, the frame importance score (FIS) calculator outputs the score sequence \(\bm{\beta} \in \mathbb{R}^{N \times 1}\). These scores, derived from comparisons between \(\{\mathbf{\check{F}}_n\}_{n=2}^{N-1}\) and \(\{\mathbf{F}_n\}_{n=2}^{N-1}\), are assigned to \(\bm{\beta}(2:N-1)\). \(\bm{\beta}(1)\) and \(\bm{\beta}(N)\) are set to sufficiently large values.
In the end, the mask generator determines which frames among the $N$ frames are key frames based on the \(\bm{\beta}\) and channel conditions (e.g., SNR), and outputs $\mathbf{v}$. 

The details of the four modules in Fig.~\ref{fig:AK_extractor} are as follows:
\subsubsection{IFNet}

IFNet diverges from traditional dual optical flow networks by adopting a coarse-to-fine strategy that directly estimates optical flow for intermediate frames. Simple linear interpolation of forward and backward flows may falter in the presence of rapid motion or complex backgrounds. In contrast, our approach mitigates artifacts and blurriness by iteratively refining optical flow estimation. The specific structure of the IFNet comprises multiple incremental flow blocks (IFBlocks). IFNet utilizes the preceding and succeeding frames of the current frame, namely $\mathbf{F}_{\text{pre}}$ and $\mathbf{F}_{\text{next}}$, to calculate the optical flow \( \mathbf{O}_{\text{cur}} \) of the current frame. IFNet employs a coarse-to-fine strategy through three IFBlocks to progressively enhance the accuracy of optical flow estimation. Each IFBlock utilizes a distinct downsampling factor $\alpha_i$ (set as 4, 2, 1 respectively) to reduce the spatial dimensions of the image and extracts features using $\text{convolutional 2D}$ layers, denoted as $\text{Conv2d}$. As resolution is incrementally enhanced, finer details are captured. Finally, upsampling restores the image to its original resolution through $\text{deconvlutional 2d}$ layers, coupled with Bilinear Resize operations, achieving high-precision optical flow estimation.

This coarse-to-fine approach iteratively corrects motion estimation inaccuracies. By refining optical flow across stages, our method effectively reduces artifacts and blurriness, resulting in more precise and higher-quality interpolation of intermediate frames.
\subsubsection{FusionNet}
FusionNet utilizes two ContextNets to extract background features from the video frames \(\mathbf{F}_{\text{pre}}\) and \(\mathbf{F}_{\text{next}}\), and considers the results of optical flow estimation through WarpBlock operations, ultimately obtaining \(\mathbf{O}^{1,2,3,4}_{\text{pre}}\) and \(\mathbf{O}^{1,2,3,4}_{\text{next}}\), respectively. ContextNet consists of four WarpBlocks. The first WarpBlock accepts the optical flow results generated by IFNet, along with either the previous frame \(\mathbf{F}_{\text{pre}}\) or the next frame \(\mathbf{F}_{\text{next}}\) as input. Subsequent WarpBlocks, specifically the $i$-th WarpBlock with $i=2,3,4$, take as inputs the intermediate results \(\mathbf{O}^{i-1}\) and \(\mathbf{F}^{i-1}\), which are produced by the $(i-1)$-th WarpBlock. The $i$-th WarpBlock utilizes $\text{Conv2d}$ to extract features from $\mathbf{O}^{\mathit{i}-1}$, and employs Bilinear Resize to restore the size of $\mathbf{O}^{\mathit{i}}$ to obtain $\mathbf{O}^{\mathit{i}}$. Concurrently, $\mathbf{O}^{\mathit{i}}$ is warped with $\mathbf{F}^{\mathit{i}-1}$ to yield $\mathbf{F}^{\mathit{i}}$. Each WarpBlock allows for the learning of background information to varying degrees. Then a U-Net structured network is employed to continually perform Conv2d and supplement \( \mathbf{O}^i \) to extract multi-scale features, and uses Deconv2d to perceive and reconstruct these features, ultimately yielding the frame reconstruction result.

\subsubsection{Frame Importance Score Calculator}
To assess the importance of frames based on video content changes and motion information, we develop a new evaluative metric called the frame importance score, denoted by \(\bm{\beta}\), which  can reflect the degree of difference between the reconstructed frames \(\{\mathbf{\check{F}}_n\}_{n=2}^{N-1}\) and the original frames \(\{\mathbf{F}_n\}_{n=2}^{N-1}\). A higher value in $\bm{\beta}$ indicates a greater distinction between the two images, signifying that the frame is more challenging to reconstruct, and thus, this frame is deemed more significant. The score of the $n$-th frame can be given by:
\begin{align}
\bm{\beta}(n) = f_{\text{is}}(\mathbf{\check{F}}_{n},\mathbf{F}_{n}),
\label{eq:FIS}
\end{align}
where \( f_{\text{is}}(\cdot , \cdot ) \) represents the assessment function, which is contingent upon the specific application context. 
For instance, one can calculate the Structural Similarity (SSIM)~\cite{SSIM} or the Learned Perceptual Image Patch Similarity (LPIPS)~\cite{LPIPS} between \(\mathbf{\check{F}}_{n}\) and \(\mathbf{F}_{n}\).

\subsubsection{Adaptive Frame Selection}

Before selecting key frames, it is essential to determine the key frames ratio \( \rho \).
When the channel condition is poor, it is essential to reduct transmission resources for retransmitting the most critical key frames, specifically those with higher \(\bm{\beta}\) values. This adjustment effectively mitigates the \emph{cliff-effect}, which deteriorates video quality significantly under extremely poor channel conditions.
By conserving bandwidth through prioritizing critical frames, the strategy ensures reliable reception of essential information even in challenging channel conditions.

Here, we adopt SNR, denoted by   $\gamma$, to represent channel conditions because of its simplicity and ease of acquisition in practice. We establish a quantitative relationship between $\gamma$ and \( \rho \) as follows:
\begin{align}
\rho = \sum_{i=0}^{I} a_i \cdot (\log(\gamma))^i,
\label{eq:rho}
\end{align}
where \( \rho \) represents the predicted key-frame ratio based on $\gamma$, \( a_i \) are the coefficients of the polynomial regression model, and \( I \) is the degree of the polynomial.

The polynomial coefficients \( \mathbf{a} = [a_0, a_1, \ldots, a_\text{I}] \) can be determined by minimizing the objective function \( Err \), which measures the discrepancy between observed and model-predicted key-frame ratios:
\begin{align}
\mathbf{a} = \arg \min_{\mathbf{a}} \left\{ Err + \upsilon \cdot R(\mathbf{a}) \right\}.
\end{align}
Here, \( \upsilon \) is a custom parameter that adjusts the balance between the data fit and the regularization term, and \( R(\mathbf{a}) \) is the regularization function given by:
\begin{align}
R(\mathbf{a}) =
\begin{cases}
\|\mathbf{a}\|_1 = \sum_{i=0}^{I} |a_i|, & \text{for L1 regularization}, \\
\|\mathbf{a}\|_2^2 = \sum_{i=0}^{I} a_i^2, & \text{for L2 regularization}.
\end{cases}
\end{align}
Here, \( R(\mathbf{a}) \) applies either L1 or L2 regularization to the coefficients \( \mathbf{a} \), with L1 promoting sparsity and L2 encouraging smaller values across the coefficients. \( Err \) is defined as:
\begin{align}
Err = \left( \rho - \left( \sum_{i=0}^{I} a_i \cdot (\log(\gamma))^i \right) \right)^2,
\end{align}
which quantifies the squared difference between the actual key-frame ratio and that predicted by the model.

Given $\gamma$, we can obtain $\rho$ and then $M$. Based on $M$ and $\bm{\beta}$, we can determine the key-frames by selecting the frames corresponding to the top \( M \) values in $\bm{\beta}$. These frames are then marked as key-frames by setting their positions in the sequence of \( N \) frames to 1.
\subsection{Adaptive Key-frame Interpolator}

\begin{figure}[!t]
\centering
\includegraphics[width=0.45\textwidth]{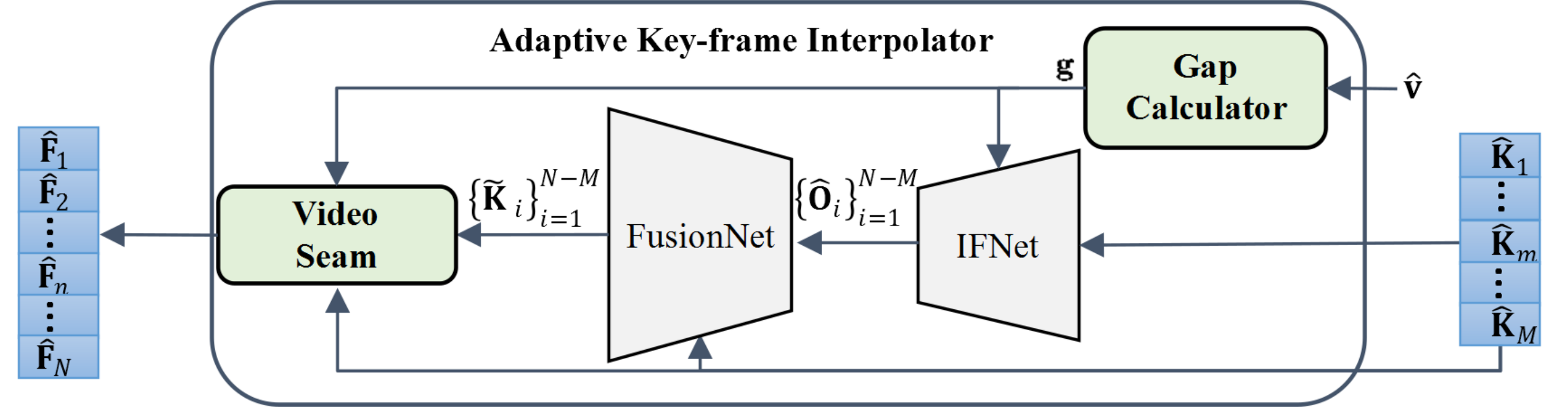}
\vspace{-5pt}
\caption{The structure of the adaptive key-frame interpolator.}
\label{fig:AK_interpolator}
\vspace{-15pt}
\end{figure}

The adaptive key-frame interpolator is designed with a simpler structure than the transmitter's adaptive key-frame extractor to accommodate the practical conditions of wireless communication. As illustrated in Fig.~\ref{fig:AK_interpolator}, its inputs are \( \{\hat{\mathbf{K}}_m\}_{m=1}^{M} \) and \( \hat{\mathbf{v}} \), which are the estimates of \( \{\mathbf{K}_m\}_{m=1}^{M} \) and \( \mathbf{v} \), repectively. Initially, the gap calculator generates a sequence \( \mathbf{g}\in\mathbb{Z}^{(M-1) \times 1}\), which records the number of non-key frames between each pair of consecutive key frames based on \( \hat{\mathbf{v}} \). Subsequently, all key frames along with \( \mathbf{g} \) are input into the IFNet to calculate the optical flow \( \{\hat{\mathbf{O}}_i\}_{i=1}^{N-M} \) of the  $(N-M)$ non-key frames interspersed between the key frames. The FusionNet then reconstructs the non-key frames \( \{ \tilde{\mathbf{K}}_i \}_{i=1}^{N-M} \), by utilizing the information from \( \{ \hat{\mathbf{O}}_i \}_{i=1}^{N-M} \). Finally, we use the \text{video seam} module to stitch together \(\{ \tilde{\mathbf{K}}_i \}_{i=1}^{N-M}\) and \(\{ \hat{\mathbf{K}}_m \}_{m=1}^{M}\) according to \( \mathbf{g} \), reconstructing the sequence of video frames \(\{ \hat{\mathbf{F}}_n \}_{n=1}^{N}\).

\section{The Proposed MSVQ}\label{section4}

To optimize compression efficiency for each key-frame, we propose MSVQ that can reduce intra-frame redundancies. MSVQ module consists of two key components: the semantic vector quantization encoder at the transmitter and the semantic vector quantization decoder at the receiver. The former possesses exceptional image compression capability, encoding the frames into multi-channel indices, while the latter accurately reconstructs the original frames from these indices.

\subsection{Semantic Vector Quantization Encoder}\label{sec:VQ-encoder}
\subsubsection{Procedure}\label{subsec:vqencoder_process}
\begin{figure}[!t]
\centering
\includegraphics[width=0.39\textwidth]{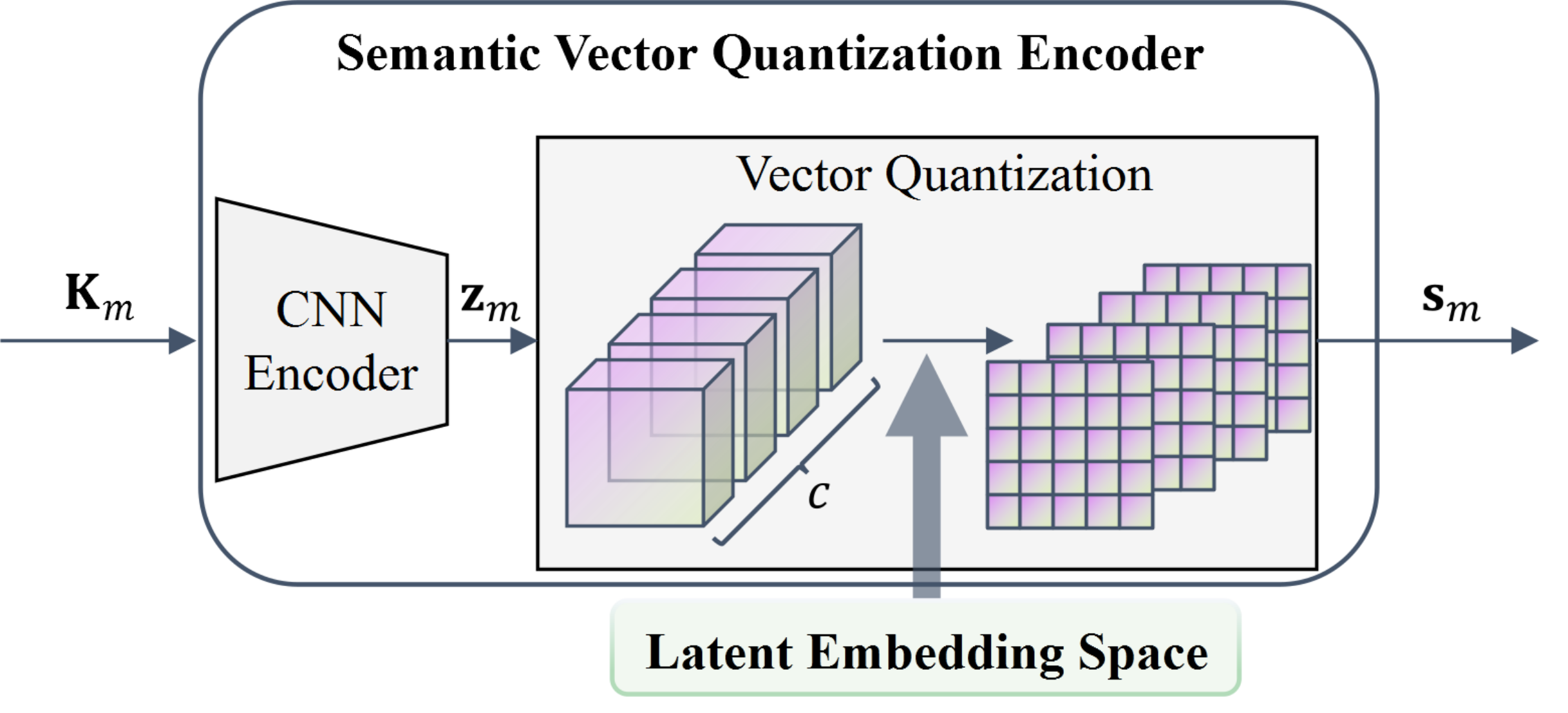}
\vspace{-10pt}
\caption{The structure of semantic quantization vector encoder.}
\label{fig:VQ-Encoder}
\vspace{-5pt}
\end{figure}

As shown in Fig.~\ref{fig:VQ-Encoder}, the encoder first applies a CNN encoder to extract multi-channel features $\mathbf{z}_m= [\mathbf{z}_m^{(1)},\ldots,\mathbf{z}_m^{(c)}] \in\mathbb{R}^{h\times w\times d\times c}$ from the $m$-th key-frame $\mathbf{K}_m$ where $\mathbf{z}_m^{(j)}\in\mathbb{R}^{h\times w\times d}$ denotes the $j$-th channel feature of the $m$-th frame, $j=1,2,\dots ,c$, and $c$ represents the number of channels. 

The feature $\mathbf{z}_m^{(j)}$ is reshaped to $\mathbf{\bar{z}}_m^{(j)} \in \mathbb{R}^{U \times d}$, where $U = hw$ and its $u$-th row, i.e., $[\mathbf{\bar{z}}_m^{(j)}]_{u,:}$, is then quantized to the index of the nearest vector in the latent embedding space $\mathbf{E}$, i.e.,
\begin{align}
\mathbf{\bar{s}}_m^{(j)}(u) = \underset{l}{\mathrm{arg}\min} \|[\mathbf{\bar{z}}_m^{(j)}]_{u,:} - \mathbf{e}_l\|_2, \label{arg}
\end{align}
where $l = 1, \ldots, L$ and \(\mathbf{\bar{s}}_m^{(j)} \in \mathbb{Z}^{U \times 1}\). Once $L$ is given, each index value, i.e., $\mathbf{\bar{s}}_m^{(j)}(u)$, can be represented by \(B = \log_2 L\) bits.

Subsequently, we concatenate these index sequences to form \(\mathbf{s}_m = [\mathbf{\bar{s}}_m^{(1)}; \ldots; \mathbf{\bar{s}}_m^{(c)}] \in \mathbb{Z}^{L_\mathbf{s} \times 1}\), where $L_\mathbf{s}=Uc$. Thus, employing the semantic vector quantization encoder compresses the data of a video frame into a vector of length \(L_\mathbf{s}\), equivalent to a bitstream of length \(L_\mathbf{s}B\).

\subsubsection{Details of CNN Encoder}
The CNN encoder in MSVQ is capable of encoding the input into a multi-channel vector rather than a single channel. The advantages of multi-channel representation over single-channel representation include increased information capacity, enhanced feature combinations, and improved image generation quality~\cite{MOC-RVQ}. 

Our CNN encoder incorporates multi-level downsampling modules to extract multi-scale features and expand the receptive field of each feature point, thus improving the model's ability to capture global structure and semantic information from the input data. Each down-sample module consists of two residual blocks, which help mitigate the vanishing gradient problem, facilitate the training of deeper networks, and improve overall network performance. To assist the network in better understanding more abstract and complex patterns that may exist in the deep features, an attention mechanism module is added in the last down-sample module. 
Finally, the features are mapped to the latent embedding space through an intermediate layer and various operations.

\subsection{Semantic Vector Quantization Decoder}
\subsubsection{Procedure}
\begin{figure}[!t]
\centering
\includegraphics[width=0.39\textwidth]{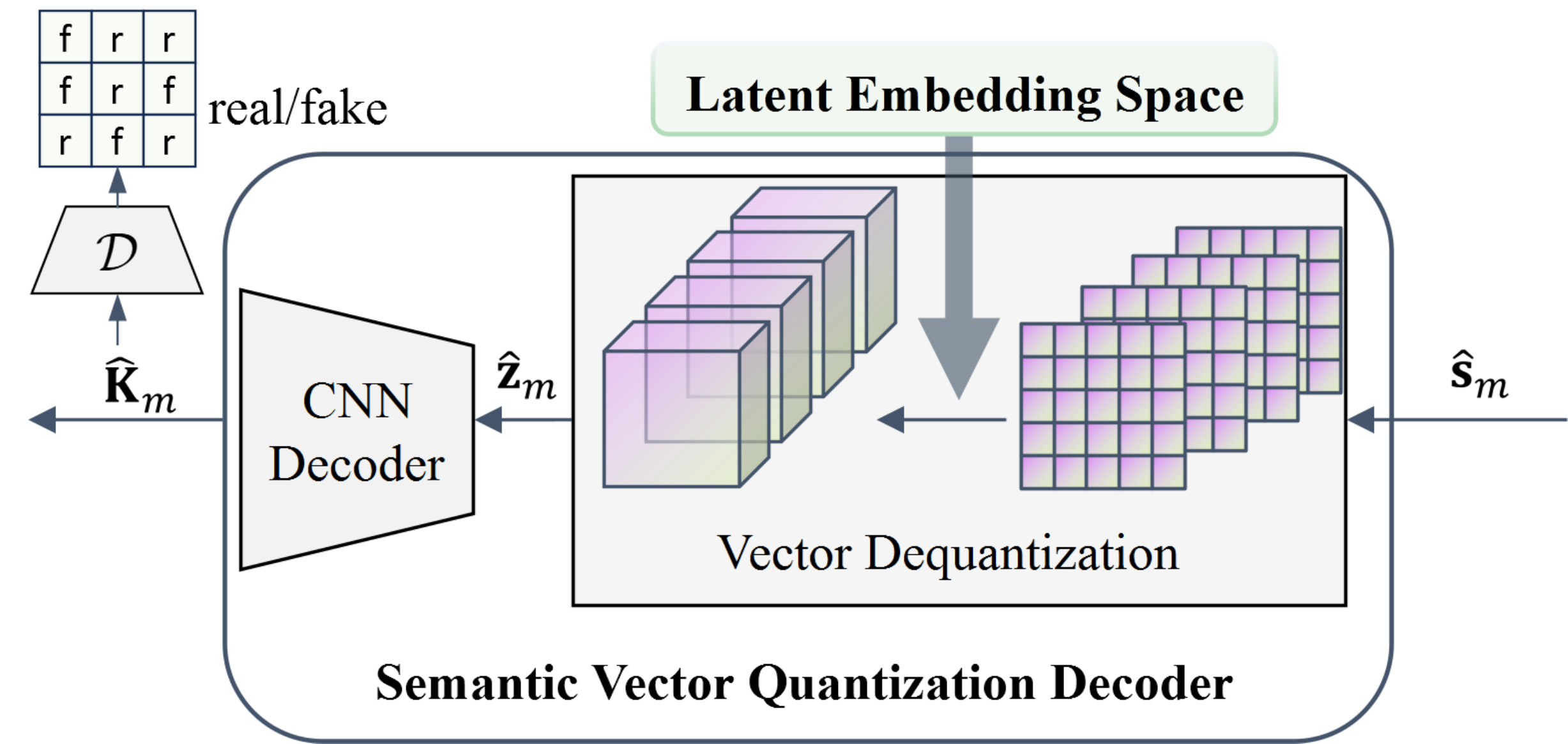}
\vspace{-10pt}
\caption{The structure of semantic vector quantization decoder.}
\label{fig:VQ-Decoder}
\vspace{-5pt}
\end{figure}

The semantic vector quantization decoder depicted in Fig.~\ref{fig:VQ-Decoder} takes the index sequence $\hat{\mathbf{s}}_m$ as input. This sequence is first mapped back to the corresponding vector $\hat{\mathbf{z}}_m$ in the latent embedding space. Subsequently, it is decoded using a CNN decoder to reconstruct the key-frame $\mathbf{K}_m$. Finally, a patch-based discriminator $\mathcal{D}$ is employed to assess the constructed frame $\hat{\mathbf{K}}_m$. It should be pointed out that during the training phase of this model, we directly assign $\hat{\mathbf{s}}_m = \mathbf{s}_m$.

\subsubsection{Details of CNN Decoder and Discriminator}
Compared to the CNN encoder, the CNN decoder in MSVQ exhibits a mirrored architecture in certain aspects. The decoder includes an intermediate layer, multiple up-sample modules, and concludes with a final Norm, SiLU, and Conv2d. The structures of the residual block and the attention mechanism module are consistent with those in the CNN encoder.

However, the CNN decoder is not completely symmetrical to the CNN encoder. It is noteworthy that all Norms in the CNN decoder, including residual blocks and attention module blocks, are replaced by spatial conditional normalization (SCN). This ensures that the same quantized indices produce results that are not identical but more natural at different positions, thereby reducing artifacts and discontinuities in the generated frames compared to conventional DL-based VQ methods~\cite{fu_vector_2023}. 

The SCN module initially standardizes the intermediate feature map $\mathbf{f}^{i} \in \mathbb{R}^{C_{\mathbf{f}^{i}} \times H_{\mathbf{f}^{i}} \times W_{\mathbf{f}^{i}}}$ via group normalization to eliminate stylistic variations, where \( C_{\mathbf{f}^{i}} \) represents the number of channels, while \( H_{\mathbf{f}^{i}} \) and \( W_{\mathbf{f}^{i}} \) represent the height and width of the feature map respectively. 
Then, it integrates the embedded vectors $\hat{\mathbf{z}}_m$ as auxiliary information into the input feature maps. Specifically, this is achieved using \eqref{SpatialNorm}:

\begin{align}
\mathbf{f}^{i+1} = \frac{\mathbf{f}^{i} - \mu_{\text{GN}}\left(\mathbf{f}^{i}\right)}{\sigma_{\text{GN}}\left(\mathbf{f}^{i}\right)} \odot \Theta_y\left(\Phi\left(\hat{\mathbf{z}}_m\right)\right) + \Theta_b\left(\Phi\left(\hat{\mathbf{z}}_m\right)\right), \label{SpatialNorm}
\end{align}
where function $\Phi(\cdot)$ interpolates and adjusts the auxiliary input $\hat{\mathbf{z}}_m$ to match the size of the input feature map $\mathbf{f}^{i}$, $\Phi\left(\hat{\mathbf{z}}_m\right)\in \mathbb{R}^{ c \times H_{\mathbf{f}^{i}} \times W_{\mathbf{f}^{i}}}$ and $c$ represents the number of channels described in Section \ref{sec:VQ-encoder}. $\Theta_y$ and $\Theta_b$ are two learnable affine transformations that map the adjusted $\hat{\mathbf{z}}_m$ to the same number of channels as $\mathbf{f}^{i}$. Additionally, $\mu_{\text{GN}}$ and $\sigma_{\text{GN}}$ respectively represent the mean and standard deviation of the group normalization applied to $\mathbf{f}^{i}$.
In summary, the normalized feature map is multiplied and added with the convolutionally mapped auxiliary information to generate the final output feature map $\mathbf{f}^{i+1}$.

\subsection{Loss Function}
During the training phase of the encoder, the loss function we use consists of three components: reconstruction loss, vector quantization loss, and adversarial loss. 

The reconstruction loss measures the difference between the frames generated by the model and the original ones, which is given by: 
\begin{align}
\mathcal{L}_{\text{rec}}(\mathbf{K}_m,\hat{\mathbf{K}}_m) = \|\mathbf{K}_m - \mathbf{\hat{K}}_m\|_2^2 + \lambda \mathcal{L}_{\text{perceptual}}(\mathbf{K}_m, \hat{\mathbf{K}}_m), \label{rec_loss}
\end{align}
where $\lambda$ is the weight coefficient.
$\mathcal{L}_{\text{perceptual}}$ quantifies the perceptual distance between the generated frames and the original frames by computing the distance between the feature maps extracted from the generated and input frames through networks. In this paper, the feature outputs of five convolutional layers in the VGG-16 model \cite{VGG16} are selected to compute the perceptual loss as:
\begin{align}\mathcal{L}_{\text{perceptual}}(\mathbf{K}_m,\hat{\mathbf{K}}_m)=\sum_{j=0}^{4} \frac1{C_jH_jW_j}\parallel\varphi_j(\mathbf{K}_m)-\varphi_j(\hat{\mathbf{K}}_m)\parallel_2^2,
\label{per}
\end{align}
where $\varphi_j$ is the output of the $j$-th layer of the feature extraction network, and $C_j$, $H_j$, $W_j$ represent the number of channels, height, and width of the feature maps at the $j$-th layer, respectively.

The vector quantization loss is incurred during the quantization process due to the straight-through gradient estimator method, which can be calculated as:
\begin{align}\mathcal{L}_{\text{VQ}}(\mathbf{z}_m,\hat{\mathbf{z}}_m) =\|\mathrm{sg}[\mathbf{z}_m]-\hat{\mathbf{z}}_m\|_2^2+\|\mathrm{sg}[\hat{\mathbf{z}}_m]-\mathbf{z}_m\|_2^2.\label{vq_loss}\end{align}
Here $\mathrm{sg}[ \cdot  ]$ denotes the stop-gradient operation.

The adversarial loss is given by: 
\begin{align}\mathcal{L}_{\mathrm{GAN}}(\mathbf{K}_m,\hat{\mathbf{K}}_m;\mathbf{\theta})=\log \mathcal{D}(\mathbf{K}_m;\mathbf{\theta})+\log(1-\mathcal{D}(\hat{\mathbf{K}}_m;\mathbf{\theta})), \label{GAN_loss}\end{align}
where $\mathbf{\theta}$ represents learnable parameters in $\mathcal{D}$.

The whole objective function for identifying the optimal model can be expressed as:
\begin{align}
\mathcal{Q}^* &= \underset{\mathbf{w}_\mathsf{E}, \mathbf{w}_\mathsf{G}, \mathbf{w}_\mathcal{Z}}{\operatorname*{\arg\min}} \underset{\mathbf{\theta}}{\operatorname*{\max}} \mathbb{E}_{\mathbf{K}_m \sim p(\mathbf{K}_m)} \Big[ \mathcal{L}_{\text{rec}} + \mu_1 \mathcal{L}_{\text{VQ}} + \mu_2 \mathcal{L}_{\text{GAN}} \Big],
\end{align}
where \(\mathbf{w}_\mathsf{E}\), \(\mathbf{w}_\mathsf{G}\), and \(\mathbf{w}_\mathcal{Z}\) denote the learnable parameters of the CNN encoder, the CNN decoder, and the latent embedding space, respectively. \(\mu_1\) and \(\mu_2\) are weighting coefficients that adjust the contributions of the vector quantization loss and the generative adversarial network loss to the total loss. \(\mathbb{E}_{\mathbf{K}_m \sim p(\mathbf{K}_m)}\) represents the expected value over the distribution \(p(\mathbf{K}_m)\) of the input frame \(\mathbf{K}_m\).

\section{Adjustable Index Selector and Restorer}\label{section5}
Using the semantic vector quantization encoder, the $m$-th key frame is represented by the sequence of indices \(\mathbf{s}_m\). 
Letting \(\mathbf{q}_m \in \mathbb{R}^{L_\mathbf{s} \times d}\) represent the matrix that consists of $L_\mathbf{s}$ quantized feature vectors for the $m$-th key frame, we have 
\begin{align}
[\mathbf{q}_m]_{i,:} = \mathbf{e}_{\mathbf{s}_m(i)}^T \label{eq_s2q}, 
\end{align}
where \(i = 1, 2, \ldots, L_\mathbf{s}\) and $\mathbf{e}_{\mathbf{s}_m(i)}$ refers to the $\mathbf{s}_m(i)$-th basis vector in the latent embedding space \(\mathbf{E}\).

In practice, the quantized feature vectors between adjacent key frames may exhibit similarity. By computing the similarity between the feature vectors of the \(\mathbf{K}_{m-1}\) and \(\mathbf{K}_m\), we can assess the redundancy between consecutive frames.

Take cosine similarity as an example, which is given by:
\begin{align}
\mathit{sim} = \frac{\langle[\mathbf{q}_{m-1}]_{i,:}, [\mathbf{q}_m]_{i,:}\rangle}{\|[\mathbf{q}_{m-1}]_{i,:}\|_2 \cdot \|[\mathbf{q}_m]_{i,:}\|_2}.\label{sim}
\end{align}
If $sim$ exceeds a predefined threshold \(\eta\), it indicates substantial redundancy between the two frames. In such cases, transmission can be optimized by selectively transmitting only the indices that exhibit significant changes, thereby reducing the volume of data transmitted while minimizing the impact on video reconstruction quality. 

Let \(\mathbf{s}_\eta\) denote the final sequence of selected indices and define:
\begin{align}
\mathbf{p} = [\mathbf{p}_2; \ldots; \mathbf{p}_M] \label{eq_position},
\end{align}
where $\mathbf{p}_m$ is a length $L_\mathbf{s}$ vector whose element is 0 or 1, and \(m = 2, \ldots, M\). Specifically, if the cosine similarity between \([\mathbf{q}_{m-1}]_{i,:}\) and \([\mathbf{q}_m]_{i,:}\) is less than \(\eta\), we set \(\mathbf{p}_m(i)=1\) and include \(\mathbf{s}_m(i)\) in \(\mathbf{s}_\eta\). Otherwise, we set \(\mathbf{p}_m(i)=0\) and keep \(\mathbf{s}_\eta\) unchanged.

Upon receiving the estimates of  \(\mathbf{s}_\eta\) and \(\mathbf{p}\), i.e., \(\hat{\mathbf{s}}_\eta\) and \(\hat{\mathbf{p}}\) at the receiver, the adjustable index restorer reconstructs the key frame index sequence \(\{\hat{\mathbf{s}}_m\}_{m=1}^{M}\). Initially, $\hat{\mathbf{s}}_1$ is initialized using $\hat{\mathbf{s}}_\eta$.
Then for each subsequent key frame $m = 2,\ldots,M$, $\hat{\mathbf{s}}_m$ is reconstructed based on $\hat{\mathbf{s}}_{m-1}$ and using $\hat{\mathbf{s}}_\eta$ with consideration of $\hat{\mathbf{p}}$. 

As shown in Fig.~\ref{fig:whole_figure}, after processing the original video frames \(\{\mathbf{F}_n\}_{n=1}^{N}\) through the adaptive key-frame extractor, semantic vector quantization encoder, and adjustable index selector modules, we can finally obtain \(\mathbf{v}\), \(\mathbf{s}_\eta\), and \(\mathbf{p}\). Define the \emph{bit compression ratio} (BCR) as the ratio of the number of bits required for the final transmission sequence to the original video frames. Thus, the BCR of the proposed VQ-DeepVSC is given by:
\begin{align}
\text{BCR} = \frac{N + L_{\eta} \times B + L_\mathbf{s} \times (M-1)}{C_\text{F} \times H_\text{F} \times W_\text{F} \times 8 \times N}, \label{all_cr}
\end{align}
where \(L_{\eta}\) is the length of \(\mathbf{s}_\eta\), and $N$, $L_{\eta}B$, and $L_\mathbf{s}(M-1)$ are the number of bits required for \(\mathbf{v}\), \(\mathbf{s}_\eta\), and \(\mathbf{p}\), respectively.

\section{Experiments}\label{section6}
In this section, we provide a detailed description of the training configurations for the AKEI and MSVQ stages, as well as the conditions for the transmission experiments. We focus on the significance of AKEI and the adjustable index selector and restorer and their impact on system performance. Finally, we compare VQ-DeepVSC with the traditional H.265 method through experiments.

\subsection{Implementation Details}\label{Implementation_Details}
In simulations, we employ low-density parity-check code (LDPC) codes with a block length of 648 bits and a 3/4 rate to encode $\mathbf{b}$. Both quadrature phase shift keying (QPSK) and 16-quadrature amplitude modulation (16QAM)  signaling schemes are utilized.

The subsequent experimental evaluation assesses the performance of the proposed VQ-DeepVSC using the UCF101~\cite{soomro2012ucf101} dataset. The UCF101 dataset, comprising 101 distinct video categories, allows for a thorough examination of the system's  capabilities. From each category, we randomly select a subset of 10 videos, creating a test dataset of 1,010 videos. This diverse test set allows for a rigorous evaluation of the efficacy and reliability of VQ-DeepVSC in video transmission, ensuring a more robust assessment of the performance of the different modules within the system.

To evaluate the system performance, we employ MS-SSIM~\cite{MSSSIM} to compare the multiscale structural similarity at the pixel level between each frame of the original video and its reconstructed counterpart, and utilize LPIPS~\cite{LPIPS} to assess the perceptual quality of the reconstructed video from the human perspective.

\begin{figure*}[!htbp]
\centering
\includegraphics[width=0.7\textwidth]{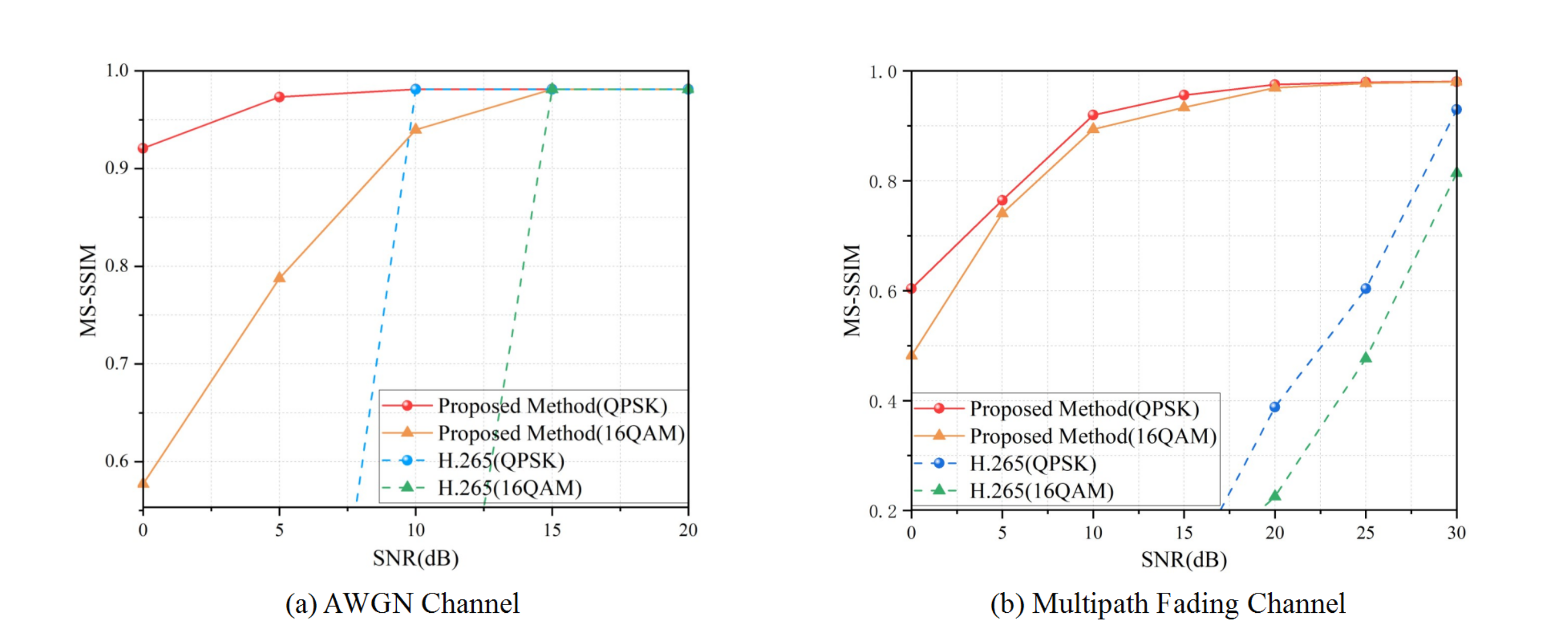}
\vspace{-10pt}
\caption{MS-SSIM comparisons between the proposed VQ-DeepVSC ($\text{BCR}=0.023$) and H.265 ($\text{BCR}=0.024$) under different channels.}
\label{MS-SSIM_di_channel}
\vspace{-15pt}
\end{figure*}

\begin{figure*}[!htbp]
\centering
\includegraphics[width=0.7\textwidth]{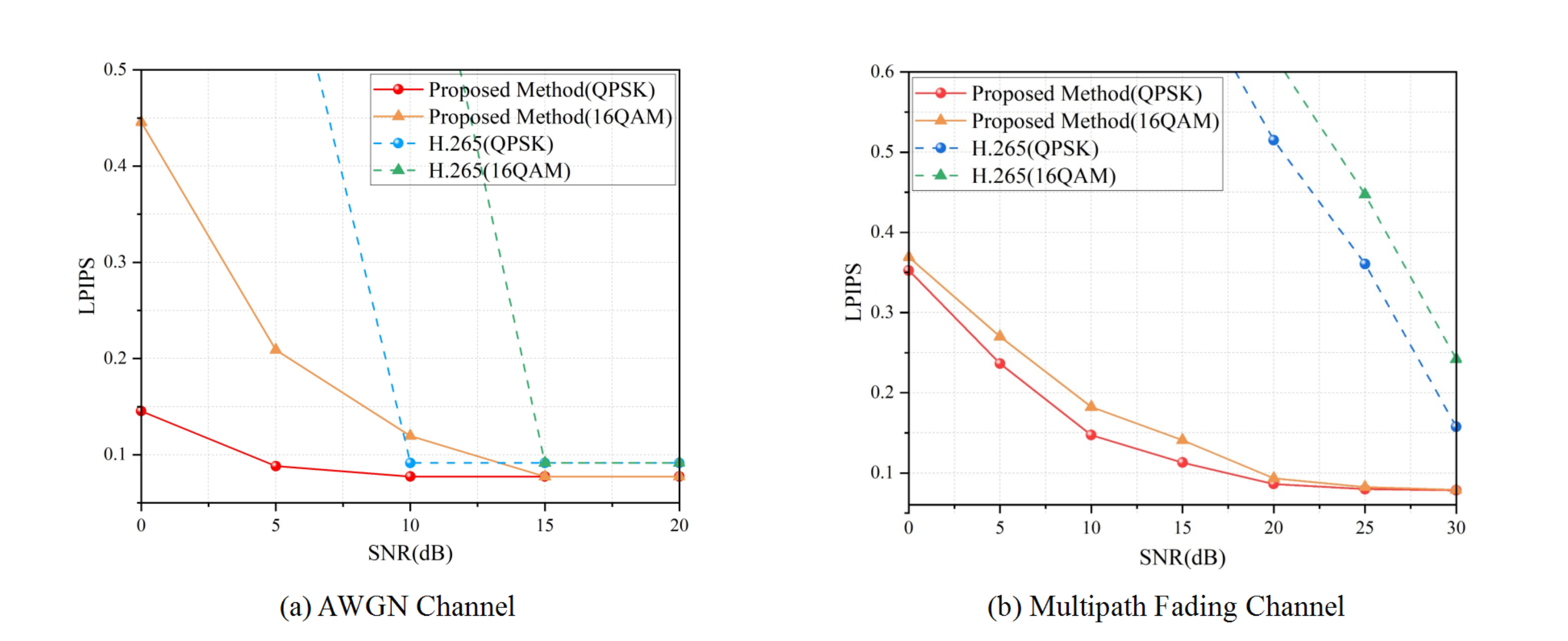}
\vspace{-10pt}
\caption{LPIPS comparisons between the proposed VQ-DeepVSC ($\text{BCR}=0.023$) and H.265 ($\text{BCR}=0.024$) under different channels.}
\label{LPIPS_di_channel}
\vspace{-15pt}
\end{figure*}

\subsection{Comparative Analysis of Methods}

We compare the performance of VQ-DeepVSC with typical and commonly used H.265 video compression codecs for video transmission. We utilize the FFMPEG~\cite{ffmpeg} library to perform H.265 encoding and decoding. The BCR of H.265 is given by:
\begin{align}\text{BCR} = \frac{B_s\times 8}{C_\text{F} \times H_\text{F} \times W_\text{F} \times 8 \times N}, \label{cr_h265}\end{align}
where $B_s$ represents the file size in bytes after H.265 compression. 
According to (\ref{cr_h265}), the BCR for H.265 on the test dataset is 0.024.
For VQ-DeepVSC, we set \(\eta\) to 1, and based on~(\ref{all_cr}), we can obtain the BCR of our system is 0.023, which is slightly lower than that of H.265.

Fig. \ref{MS-SSIM_di_channel} and Fig. \ref{LPIPS_di_channel} show the comparisons between the proposed VQ-DeepVSC and H.265 video codecs under AWGN and multipath fading channel, respectively. 
From the figures, it can be observed that the proposed VQ-DeepVSC outperforms H.265 in terms of both MS-SSIM and LPIPS, especially at medium to low SNRs. 
This is due to H.265 being sensitive to noise, thereby limiting its effectiveness at medium to low SNRs. In contrast, our method not only achieves reliable video transmission at low SNRs but also maintains high video quality. 
It can also be observed that under multipath fading channel, H.265 fails to transmit videos effectively, whereas the proposed method maintains robust video transmission and ensures high quality. 

\section{Conclusions}\label{section7}
In this paper, we present the VQ-DeepVSC system, which employs an innovative dual-stage vector quantization framework to enhance video transmission quality and efficiency over wireless channels. This system effectively mitigates the \emph{cliff-effect} and maintains significant data compression even under low SNR conditions, ensuring robust video transmission.
The proposed VQ-DeepVSC system employs the AKEI module in the first stage to extract key frames, reducing inter-frame redundancy and enhancing transmission robustness under low SNR conditions.
In the second stage, the MSVQ module compresses these key frames using a shared latent embedding space, effectively reducing intra-frame redundancy and supporting high-resolution video transmission. Additionally, the adjustable index selector and restorer further reduce inter-frame redundancy by compressing key-frame indices, enabling more flexible compression rates.
Experimental results demonstrate that the proposed VQ-DeepVSC achieves higher compression degree and higher video quality than the H.265 standard, especially under low SNRs or multi-path channels. Simulation results also demonstrate that the proposed VQ-DeepVSC has excellent generalization capabilities, making it suitable for various types of video transmission.

\bibliographystyle{ieeetr}
\bibliography{reference.bib}

\begin{thebibliography}{10}

\bibitem{al_fuqaha_internet_2015}
Ala Al-Fuqaha, Mohsen Guizani, Mehdi Mohammadi, Mohammed Aledhari, and Moussa Ayyash.
\newblock Internet of {Things}: {A} {Survey} on {Enabling} {Technologies}, {Protocols}, and {Applications}.
\newblock {\em IEEE Communications Surveys \& Tutorials}, 17(4):2347--2376, 2015.

\bibitem{zhang_6g_2019}
Zhengquan Zhang, Yue Xiao, Zheng Ma, Ming Xiao, Zhiguo Ding, Xianfu Lei, George~K. Karagiannidis, and Pingzhi Fan.
\newblock {6G} {Wireless} {Networks}: {Vision}, {Requirements}, {Architecture}, and {Key} {Technologies}.
\newblock {\em IEEE Vehicular Technology Magazine}, 14(3):28--41, September 2019.

\bibitem{wiegand_overview_2003}
T.~Wiegand, G.J. Sullivan, G.~Bjontegaard, and A.~Luthra.
\newblock Overview of the {H}.264/{AVC} video coding standard.
\newblock {\em IEEE Transactions on Circuits and Systems for Video Technology}, 13(7):560--576, July 2003.

\bibitem{sullivan_overview_2012}
Gary~J. Sullivan, Jens-Rainer Ohm, Woo-Jin Han, and Thomas Wiegand.
\newblock Overview of the {High} {Efficiency} {Video} {Coding} ({HEVC}) {Standard}.
\newblock {\em IEEE Transactions on Circuits and Systems for Video Technology}, 22(12):1649--1668, December 2012.

\bibitem{shi_semantic_2021}
Guangming Shi, Yong Xiao, Yingyu Li, and Xuemei Xie.
\newblock From {Semantic} {Communication} to {Semantic}-{Aware} {Networking}: {Model}, {Architecture}, and {Open} {Problems}.
\newblock {\em IEEE Communications Magazine}, 59(8):44--50, August 2021.

\bibitem{luo_semantic_2022}
Xuewen Luo, Hsiao-Hwa Chen, and Qing Guo.
\newblock Semantic {Communications}: {Overview}, {Open} {Issues}, and {Future} {Research} {Directions}.
\newblock {\em IEEE Wireless Communications}, 29(1):210--219, February 2022.

\bibitem{yang_semantic_2023}
Wanting Yang, Hongyang Du, Zi~Qin Liew, Wei Yang~Bryan Lim, Zehui Xiong, Dusit Niyato, Xuefen Chi, Xuemin Shen, and Chunyan Miao.
\newblock Semantic {Communications} for {Future} {Internet}: {Fundamentals}, {Applications}, and {Challenges}.
\newblock {\em IEEE Communications Surveys \& Tutorials}, 25(1):213--250, 2023.

\bibitem{xu_wireless_2022}
Jialong Xu, Bo~Ai, Wei Chen, Ang Yang, Peng Sun, and Miguel Rodrigues.
\newblock Wireless {Image} {Transmission} {Using} {Deep} {Source} {Channel} {Coding} {With} {Attention} {Modules}.
\newblock {\em IEEE Transactions on Circuits and Systems for Video Technology}, 32(4):2315--2328, April 2022.

\bibitem{hu_robust_2022}
Qiyu Hu, Guangyi Zhang, Zhijin Qin, Yunlong Cai, Guanding Yu, and Geoffrey~Ye Li.
\newblock Robust {Semantic} {Communications} {Against} {Semantic} {Noise}.
\newblock In {\em 2022 {IEEE} 96th {Vehicular} {Technology} {Conference} ({VTC2022}-{Fall})}, pages 1--6, London, United Kingdom, September 2022. IEEE.

\bibitem{zhou_semantic_2022}
Qingyang Zhou, Rongpeng Li, Zhifeng Zhao, Chenghui Peng, and Honggang Zhang.
\newblock Semantic {Communication} {With} {Adaptive} {Universal} {Transformer}.
\newblock {\em IEEE Wireless Communications Letters}, 11(3):453--457, March 2022.

\bibitem{tong_federated_2021}
Haonan Tong, Zhaohui Yang, Sihua Wang, Ye~Hu, Omid Semiari, Walid Saad, and Changchuan Yin.
\newblock Federated {Learning} for {Audio} {Semantic} {Communication}.
\newblock {\em Frontiers in Communications and Networks}, 2:734402, September 2021.

\bibitem{weng_semantic_2021}
Zhenzi Weng and Zhijin Qin.
\newblock Semantic {Communication} {Systems} for {Speech} {Transmission}.
\newblock {\em IEEE Journal on Selected Areas in Communications}, 39(8):2434--2444, August 2021.

\bibitem{xu_wireless_2022-1}
Jialong Xu, Bo~Ai, Wei Chen, Ang Yang, Peng Sun, and Miguel Rodrigues.
\newblock Wireless {Image} {Transmission} {Using} {Deep} {Source} {Channel} {Coding} {With} {Attention} {Modules}.
\newblock {\em IEEE Transactions on Circuits and Systems for Video Technology}, 32(4):2315--2328, April 2022.

\bibitem{hu_robust_2023}
Qiyu Hu, Guangyi Zhang, Zhijin Qin, Yunlong Cai, Guanding Yu, and Geoffrey~Ye Li.
\newblock Robust {Semantic} {Communications} {With} {Masked} {VQ}-{VAE} {Enabled} {Codebook}.
\newblock {\em IEEE Transactions on Wireless Communications}, 22(12):8707--8722, December 2023.

\bibitem{bourtsoulatze_deep_2019}
Eirina Bourtsoulatze, David Burth~Kurka, and Deniz Gunduz.
\newblock Deep {Joint} {Source}-{Channel} {Coding} for {Wireless} {Image} {Transmission}.
\newblock {\em IEEE Transactions on Cognitive Communications and Networking}, 5(3):567--579, September 2019.

\bibitem{tung_deepwive_2022}
Tze-Yang Tung and Deniz Gunduz.
\newblock {DeepWiVe}: {Deep}-{Learning}-{Aided} {Wireless} {Video} {Transmission}.
\newblock {\em IEEE Journal on Selected Areas in Communications}, 40(9):2570--2583, September 2022.

\bibitem{ma_image_2020}
Siwei Ma, Xinfeng Zhang, Chuanmin Jia, Zhenghui Zhao, Shiqi Wang, and Shanshe Wang.
\newblock Image and {Video} {Compression} {With} {Neural} {Networks}: {A} {Review}.
\newblock {\em IEEE Transactions on Circuits and Systems for Video Technology}, 30(6):1683--1698, June 2020.

\bibitem{yang_learning_2020}
Ren Yang, Fabian Mentzer, Luc Van~Gool, and Radu Timofte.
\newblock Learning for {Video} {Compression} {With} {Hierarchical} {Quality} and {Recurrent} {Enhancement}.
\newblock In {\em 2020 {IEEE}/{CVF} {Conference} on {Computer} {Vision} and {Pattern} {Recognition} ({CVPR})}, pages 6627--6636, Seattle, WA, USA, June 2020. IEEE.

\bibitem{lu_end--end_2021}
Guo Lu, Xiaoyun Zhang, Wanli Ouyang, Li~Chen, Zhiyong Gao, and Dong Xu.
\newblock An {End}-to-{End} {Learning} {Framework} for {Video} {Compression}.
\newblock {\em IEEE Transactions on Pattern Analysis and Machine Intelligence}, 43(10):3292--3308, October 2021.

\bibitem{hu_fvc_2021}
Zhihao Hu, Guo Lu, and Dong Xu.
\newblock {FVC}: {A} {New} {Framework} towards {Deep} {Video} {Compression} in {Feature} {Space}.
\newblock In {\em 2021 {IEEE}/{CVF} {Conference} on {Computer} {Vision} and {Pattern} {Recognition} ({CVPR})}, pages 1502--1511, Nashville, TN, USA, June 2021. IEEE.

\bibitem{zhang_deep_2023}
Zhenguo Zhang, Qianqian Yang, Shibo He, and Jiming Chen.
\newblock Deep {Learning} {Enabled} {Semantic} {Communication} {Systems} for {Video} {Transmission}.
\newblock In {\em 2023 {IEEE} 98th {Vehicular} {Technology} {Conference} ({VTC2023}-{Fall})}, pages 1--5, Hong Kong, Hong Kong, October 2023. IEEE.

\bibitem{dong_demo_2022}
Hao Dong, Weijie Yue, and Kai Niu.
\newblock A {Demo} of {Semantic} {Communication}: {Rosefinch}.
\newblock In {\em 2022 14th {International} {Conference} on {Wireless} {Communications} and {Signal} {Processing} ({WCSP})}, pages 373--377, Nanjing, China, November 2022. IEEE.

\bibitem{gong_adaptive_2023}
Wentao Gong, Haonan Tong, Sihua Wang, Zhaohui Yang, Xinxin He, and Changchuan Yin.
\newblock Adaptive {Bitrate} {Video} {Semantic} {Communication} over {Wireless} {Networks}.
\newblock In {\em 2023 {International} {Conference} on {Wireless} {Communications} and {Signal} {Processing} ({WCSP})}, pages 122--127, Hangzhou, China, November 2023. IEEE.

\bibitem{niu_deep_2023}
Haiwen Niu, Luhan Wang, Zhaoming Lu, Keliang Du, and Xiangming Wen.
\newblock Deep {Learning} {Enabled} {Video} {Semantic} {Transmission} {Against} {Multi}-{Dimensional} {Noise}.
\newblock In {\em 2023 {IEEE} {Globecom} {Workshops} ({GC} {Wkshps})}, pages 1267--1272, Kuala Lumpur, Malaysia, December 2023. IEEE.

\bibitem{bao_mdvscwireless_2023}
Zhicheng Bao, Haotai Liang, Chen Dong, Xiaodong Xu, and Geng Liu.
\newblock {MDVSC}—{Wireless} {Model} {Division} {Video} {Semantic} {Communication} for {6G}.
\newblock In {\em 2023 {IEEE} {Globecom} {Workshops} ({GC} {Wkshps})}, pages 1572--1578, Kuala Lumpur, Malaysia, December 2023. IEEE.

\bibitem{liang_vista_2023}
Chengsi Liang, Xiangyi Deng, Yao Sun, Runze Cheng, Le~Xia, Dusit Niyato, and Muhammad~Ali Imran.
\newblock {VISTA}: {Video} {Transmission} over {A} {Semantic} {Communication} {Approach}.
\newblock In {\em 2023 {IEEE} {International} {Conference} on {Communications} {Workshops} ({ICC} {Workshops})}, pages 1777--1782, Rome, Italy, May 2023. IEEE.

\bibitem{jiang_wireless_2023}
Peiwen Jiang, Chao-Kai Wen, Shi Jin, and Geoffrey~Ye Li.
\newblock Wireless {Semantic} {Communications} for {Video} {Conferencing}.
\newblock {\em IEEE Journal on Selected Areas in Communications}, 41(1):230--244, January 2023.

\bibitem{liu_efficient_2023}
Jiajia Liu, Yunlong Lu, Hao Wu, and Yueyue Dai.
\newblock Efficient {Resource} {Allocation} and {Semantic} {Extraction} for {Federated} {Learning} {Empowered} {Vehicular} {Semantic} {Communication}.
\newblock In {\em 2023 {IEEE} 98th {Vehicular} {Technology} {Conference} ({VTC2023}-{Fall})}, pages 1--5, Hong Kong, Hong Kong, October 2023. IEEE.

\bibitem{wang_wireless_2023}
Sixian Wang, Jincheng Dai, Zijian Liang, Kai Niu, Zhongwei Si, Chao Dong, Xiaoqi Qin, and Ping Zhang.
\newblock Wireless {Deep} {Video} {Semantic} {Transmission}.
\newblock {\em IEEE Journal on Selected Areas in Communications}, 41(1):214--229, January 2023.

\bibitem{fu_vector_2023}
Qifan Fu, Huiqiang Xie, Zhijin Qin, Gregory Slabaugh, and Xiaoming Tao.
\newblock Vector quantized semantic communication system.
\newblock {\em IEEE Wireless Communications Letters}, 12(6):982--986, 2023.

\bibitem{VQVAE}
Aaron van~den Oord, Oriol Vinyals, and Koray Kavukcuoglu.
\newblock Neural discrete representation learning.
\newblock In {\em Proc. Advances in Neural Information Processing Systems (NeurIPS), Long Beach, Dec.}, volume~30, 2017.

\bibitem{Unet}
Olaf Ronneberger, Philipp Fischer, and Thomas Brox.
\newblock U-net: Convolutional networks for biomedical image segmentation.
\newblock In {\em Proc. International Conference on Medical Image Computing and Computer-Assisted Intervention (MICCAI), Munich, Oct.}, pages 234--241, 2015.

\bibitem{MOC-RVQ}
Yingbin Zhou, Yaping Sun, Guanying Chen, Xiaodong Xu, Hao Chen, Binhong Huang, Shuguang Cui, and Ping Zhang.
\newblock Moc-rvq: Multilevel codebook-assisted digital generative semantic communication.
\newblock {\em arXiv preprint arXiv:2401.01272}, 2024.

\bibitem{SSIM}
Zhou Wang, A.C. Bovik, H.R. Sheikh, and E.P. Simoncelli.
\newblock Image quality assessment: from error visibility to structural similarity.
\newblock {\em IEEE Transactions on Image Processing}, 13(4):600--612, 2004.

\bibitem{LPIPS}
Richard Zhang, Phillip Isola, Alexei~A. Efros, Eli Shechtman, and Oliver Wang.
\newblock The unreasonable effectiveness of deep features as a perceptual metric.
\newblock In {\em 2018 IEEE/CVF Conference on Computer Vision and Pattern Recognition}, pages 586--595, 2018.

\bibitem{VGG16}
Karen Simonyan and Andrew Zisserman.
\newblock Very deep convolutional networks for large-scale image recognition.
\newblock In Yoshua Bengio and Yann LeCun, editors, {\em 3rd International Conference on Learning Representations (ICLR 2015), San Diego, CA, USA, May 7-9, 2015, Conference Track Proceedings}, 2015.

\bibitem{soomro2012ucf101}
Khurram Soomro, Amir~Roshan Zamir, and Mubarak Shah.
\newblock Ucf101: A dataset of 101 human actions classes from videos in the wild.
\newblock {\em arXiv preprint arXiv:1212.0402}, 2012.

\bibitem{MSSSIM}
Z.~Wang, E.P. Simoncelli, and A.C. Bovik.
\newblock Multiscale structural similarity for image quality assessment.
\newblock In {\em The Thrity-Seventh Asilomar Conference on Signals, Systems \& Computers, 2003}, volume~2, pages 1398--1402 Vol.2, 2003.

\bibitem{ffmpeg}
S.~Tomar.
\newblock Converting video formats with {FFmpeg}.
\newblock {\em Linux Journal}, 2006(146):10, 2006.

\end{thebibliography}

\end{document}